\documentclass[prl,10pt,twocolumn,preprintnumbers,amsmath,amssymb,nofootinbib]{revtex4}
\usepackage{dcolumn}
\usepackage{bm}
\usepackage{mathrsfs}
\usepackage{amsmath}
\usepackage{amssymb}
\usepackage{amsfonts}
\usepackage{eucal}
\usepackage{color}
\usepackage[active]{srcltx}%

\newcommand{\ba}{\begin{array}}
\newcommand{\ea}{\end{array}}
\newcommand{\be}{\begin{equation}}
\newcommand{\ee}{\end{equation}}
\newcommand{\bea}{\begin{eqnarray}}
\newcommand{\eea}{\end{eqnarray}}
\newcommand{\bse}{\begin{subequations}}
\newcommand{\ese}{\end{subequations}}
\newcommand{\bi}{\begin{itemize}}
\newcommand{\ei}{\end{itemize}}

\definecolor{darkgreen}{rgb}{0,0.3,0}
\definecolor{darkblue}{rgb}{0,0,0.3}
\definecolor{darkred}{rgb}{0.7,0,0}

\begin{document}

%\preprint{
%           IPM-P/2012-nnn\cr}
%{\vskip .2cm}

\title{Three Theorems on Near Horizon Extremal Vanishing Horizon Geometries}

\author{S. Sadeghian$^{\dag,\natural}$, M.M. Sheikh-Jabbari$^\natural$, M.H. Vahidinia$^\natural$ and  H. Yavartanoo$^\ddag$}

%\vspace*{20mm}
%\bigskip\medskip
\affiliation{\vspace*{3mm}$^{\dag}$  \textit{Department of Physics, Alzahra University P. O. Box 19938, Tehran 91167, Iran} 
\vspace*{2mm}\\ $^\natural$    \textit{School of Physics, Institute for Research in Fundamental
Sciences (IPM), P.O.Box 19395-5531, Tehran, Iran}   
\vspace*{2mm}\\
$^\ddag$ {State Key Laboratory of Theoretical Physics, Institute of Theoretical Physics,\vspace*{-.3mm}\\
Chinese Academy of Sciences, Beijing 100190, China}}
\vfil

\setcounter{footnote}{0}

%%%%%%%%%%%%%%%%%%%%%%%%%%%%%%%%%%%%%%%%%%%%%%%%%%%%%%%%%%%%%%%%%%%%%%%%%%%%%%%%%%%%%%%%
\begin{abstract}
\noindent
EVH black holes are Extremal  black holes with Vanishing Horizon area, where vanishing of horizon area is a result of having  a vanishing one-cycle on the horizon.  We prove three theorems regarding near horizon geometry of EVH black hole solutions to generic Einstein gravity theories in diverse dimensions. These generic gravity theories are  Einstein-Maxwell-dilaton-$\Lambda$ theories, and  gauged or ungauged supergravity theories with $U(1)$ Maxwell fields. Our three theorems are: (1) The near horizon geometry of any EVH black hole has a three dimensional maximally symmetric subspace. (2) If the energy momentum tensor of the theory satisfies strong energy condition  either this 3d part is  an AdS$_3$, or the solution is a direct product of a locally 3d flat space and a $d\!-\!3$ dimensional part. (3) These results extend to the near horizon geometry of near-EVH black holes, for which the AdS$_3$ part is replaced with BTZ geometry.

\end{abstract}
%\date{\today}
%\date{October 2003}
\maketitle

%\section{Introduction}
Black holes, their classical aspects, semi-classical and thermodynamical aspects, quantum aspects and finally black holes in real world and nature, have long been four  very active areas of research. There is a common understanding that black holes are the windows to new physics, especially when the extreme gravitational effects are concerned  \cite{Black-Holes-General}. A part of analysis of classical aspects of black holes involves constructing black hole solutions to various gravity theories in diverse dimensions, studying classification and uniqueness, and geometric aspects of these solutions \cite{ER-review}.
Getting insight into the behavior of gravity theory and its solutions in diverse dimensions not only provides a new perspective into the 4d gravity, but is also  what is expected from a variety of quantum gravity theories, most notably string or M-theory.
Despite the significant effort put into  classification and uniqueness theorems \cite{Black-Holes-General}, such theorems have been mainly robustly proven for stationary, asymptotic flat  black hole solutions to 4d Einstein-Maxwell theory.

There is a special class of black holes, extremal black holes, which have been of interest both as classical solutions and as test grounds for asking questions about quantum aspects of black holes. They also seem to be a good model for fast rotating black hole candidates \cite{Extreme-Kerr-observation}.
Extremal black holes have two coincident inner and outer (Killing) horizons, have vanishing Hawking temperature \cite{Black-Holes-General, Wald} and hence do not Hawking radiate \cite{KL-review}; they have the lowest mass in the family of black holes with given conserved charges and may be viewed as ground states for more general  non-extremal black holes; moreover, all supersymmetric black holes are necessarily extremal \cite{Ferrara-review}. These all have made extremal black holes an interesting family especially when questions about thermodynamical, semi-classical and quantum aspects of black holes are concerned; e.g. see \cite{Sen-review}.

It has been shown that when  the spatial cross sections of the horizon possess sufficiently many commuting rotational
isometries, the near horizon geometry of extremal black holes generically provides us with another family of solutions with $SL(2,R)\times U(1)^n$ isometry \cite{KL-review, KL-papers}. This new family of solutions is dubbed as Near Horizon Extremal Geometry (NHEG). Since many basic thermodynamical and quantum properties of black holes are associated with the properties at the horizon, studying such near horizon geometries and the whole NHEG family would shed light on similar questions on generic black holes. We would like to note that although near horizon limit of every extremal black hole leads to a solution in the class of NHEG's, the converse is not  necessarily true; we do not know (or are not able to explicitly construct) the black hole solution corresponding to each NHEG. There are elegant uniqueness and classification theorems proved for the NHEG's \cite{KL-papers,KL-review}. These theorems are for general Einstein gravity theories in four and five dimensions and, for a restricted class of geometries with $SL(2,R)\times U(1)^{d-3}$ isometry in generic $d$ dimensional theory with the matter field satisfying strong energy condition \cite{KL-review}. Moreover, in a semi-classical analysis, the laws of NHEG mechanics has been worked out \cite{First-law,NHEG-mechanics}. These laws parallel  laws of black hole thermodynamics \cite{Black-Holes-General}.

An interesting class of extremal black holes are Extremal Vanishing Horizon area (EVH) black holes. If we denote the surface gravity of a black hole by $\kappa$ and its horizon area by $A_h$, we define EVH black holes in the following limit \cite{EVH/CFT}
\be\label{EVH-def}
\kappa,\ A_h\to 0\,,\qquad \kappa/A_h=\text{fixed}.
\ee
Although, it can be more general, in our current treatment of EVH black holes we assume that vanishing of the horizon area is a result of having a vanishing one-cycle at the horizon.
Various examples of  asymptotic flat or AdS and, stationary or static EVH black holes in generic $d\geq3$ dimensions have been identified and studied; e.g. see \cite{EVH-examples}.
Studying EVH black holes besides the GR solution building purposes,  is motivated by the fact that as observed in all of these examples, in the near horizon limit of an EVH black hole we find an AdS$_3$ throat, a locally $SO(2,2)$ invariant part of geometry.

The analysis of EVH black holes and their near horizon dynamics is in interesting and distinct from generic NHEG's: the ''no-dynamics'' statements \cite{no-dynamics, NHEG-mechanics} for perturbations around NHEG's do not apply to the EVH black holes. This is witnessed by the appearance of the AdS$_3$ factors in the near horizon limit for EVH case. In the near-horizon EVH case we expect to remain with a part of original EVH black hole dynamical perturbations, which may be associated with excitations around the AdS$_3$ part of geometry.
This AdS$_3$, and the point that this near horizon limit is a decoupling limit, prompted two of us to propose the EVH/CFT proposal for the 4d EVH black holes, stating that the low energy dynamics around an EVH black hole is described by a 2d CFT  \cite{EVH/CFT}. The EVH/CFT proposal has been extended to the other known examples of EVH black holes \cite{EVH-examples}. 

Given the theorems applying to generic extremal black holes \cite{KL-review}, which state that the NHEG have generically an AdS$_2$ (and not AdS$_3$) throat, and recalling that EVH black holes are extremal, 
one is led to the question why and how these theorems fail for the case of EVH black holes. The key to this question comes from  the very definition of EVH black holes and that they have a vanishing one-cycle at the horizon. 
Therefore, the near horizon EVH black holes do not satisfy the smoothness assumption of the generic NHEG theorems \cite{KL-review} and hence do not necessarily obey those theorems.  In this Letter we study in exactly which way 
the EVH black holes evade those theorems. We in fact prove some general theorems regarding the Near Horizon structure of EVH (NHEVH) geometries. Our theorems, as we will argue, apply to a broad class of Einstein gravity theories in  generic dimensions.
These theories include bosonic part of the gauged or unguaged supergravity theories with $U(1)$ gauge and scalar fields.
%%%%%%%%%%%%%%%%%%%%%%%%%%%%%%%%%%%%%%%%%
\vskip 1.5mm
\textbf{General EVH and NHEVH ansatz.}
Gaussian null coordinates may be defined in a neighbourhood of any null hypersurface. In particular,  in the vicinity of a Killing horizon, the metric in the Guassian null coordinates can be written as \cite{KL-papers,KL-review}
\be\label{NGC-generic}
\begin{split}
ds^2&=2drdv+2r{\mathsf f}_i(r,y)dvdy^i-r {\mathsf F}(r,{y}) dv^2\\
&+ {\mathsf h}_{ij}(r,{y})dy^idy^j,\quad\  i, j =1, 2, \cdots d-2,
\end{split}\ee
where the Killing horizon is located at $r=0$ and defined by Killing vector $N=\partial_v$. The surface gravity and horizon area of \eqref{NGC-generic} are \cite{KL-review}
\be
\kappa=\frac12{\mathsf F}(r=0,y),\;\;\; A_h=\int_{r=0} \sqrt{{\det} {\mathsf h}}\; d^{d-2}y\,.
\ee
For an EVH black hole then $\kappa \sim A_h \sim\epsilon$ where $\epsilon$ is a small parameter measuring how close to extremality (more precisely, to ``EVH-ness'') we are. We  assume vanishing of $A_h$ is due to a vanishing one-cycle at $r = 0$ and parameterize this direction by $\phi$. We also assume $\partial_\phi$ to be a Killing direction and  hence functions in the metric do not have $\phi$ dependence. It is then more convenient to decompose $y^i$ into $(x^a,\phi)$. The leading $\epsilon$ expansion of the metric functions hence take the form \cite{Long-version}
\be\begin{split}
 {\mathsf F}(r,y)&=\epsilon { { F}}^{(1)}+r  { F}(x),\cr
{\mathsf h}_{ij}dy^idy^j&={\mathsf G}(r,x) d\phi^2+2{\mathsf g}_a(r,x) d\phi dx^a+ \hat\gamma_{ab}(r,x) dx^adx^b, \nonumber
\end{split}\ee
where $F^{(1)}$ is a positive constant, $a,b=1,2,\cdots d-3$, and
\be\label{phi-x-metric-expansion}
{\mathsf g}_a=\epsilon g_a^{(1)}(x)+rg_{a}(x) ,\ {\mathsf G}=\epsilon^2 G^{(2)}(x)+{\epsilon }r{G}^{(1)}+r^{2}G(x)\,.\nonumber
\ee
As the above form clearly shows, although the components $h_{ij}$ are smooth functions of $x^a, r$, the metric $h_{ij}$ is not invertible at the horizon. As will be made explicit below, this is at the root of the main differences of the EVH and generic extremal cases reviewed in \cite{KL-review}.

Along with $\epsilon\rightarrow 0$, we take the near horizon limit,
\be\label{NH-limit}
r\rightarrow \lambda r,\;\; v\rightarrow \frac{v}{\lambda}, \;\; \phi\rightarrow \frac{\phi}{\lambda},\qquad \lambda\rightarrow 0,
\ee
to obtain
\bea
\label{genericNH}
 ds^{2}\! &=&\!-r\left( \frac{\epsilon }{\lambda }{F^{(1)}} +rF
\right) dv^{2} +2r\left( \frac{\epsilon }{\lambda }{H}^{(1)}+rH
\right) d\phi dv  \nonumber \\
&+&\left( \frac{\epsilon ^{2}}{\lambda ^{2}}{
{G}^{(2)}} +\frac{\epsilon }{\lambda }r{G}^{(1)}+r^{2}G\right) d\phi ^{2} +2drdv+ 2r f_a   dx^{a}dv\nonumber \\
&+&   2\left( \frac{\epsilon }{\lambda }{g}_{a}^{(1)}+rg_{a}\right) dx^{a}d\phi +\gamma _{ab}dx^{a}dx^{b}+{\mathcal{O}}(\lambda,\epsilon )\;.
\eea
where $\gamma _{ab}=\hat\gamma_{ab}(r=0,x)$, $H^{(1)}$ and $H$ are first terms in the near horizon expansion of ${\mathsf f}_{\phi}$. If we take this limit such that $\epsilon\ll \lambda$ we are dealing with the near-horizon EVH geometry, while taking the limit $\epsilon\sim\lambda$ correspond to near-EVH near-horizon limit.  The  $\epsilon \gg  \lambda$ case (while $\epsilon\to 0$) corresponds to far from EVH cases and we do not discuss it here. Taking near horizon $\lambda, \epsilon/\lambda \to 0$ limit of EVH black hole solutions gives
\be\label{NHEVH}\begin{split}
ds^{2} &=r^{2}\left[ -Fdv^{2}+G d\phi ^{2}+2H d\phi dv\right]  +2drdv \\
&+2r\left[ f_{a} dx^{a}dv+g_{a} dx^{a}d\phi \right] +\gamma
_{ab} dx^{a}dx^{b},
\end{split}\ee
where all undetermined coefficients are  functions of only $x^a$. That is, imposing the EVH conditions fixes the $r$ dependence of all metric coefficients while to fix their $x^a$ dependence we need equations of motion 
on above metric and matter fields coupled to gravity.
%%%%%%%%%%%%%%%%%%%%%%%%%%%%%%%%%%%%%%%%%

\vskip 1.5mm
\textbf{General implications of Einstein equations.} To restrict further the form of metric \eqref{NHEVH}  we use smoothness properties and Einstein equations which in $d$ dimensions take the form
\be\label{Ein-Eq}
R_{\mu\nu}=T_{\mu\nu}-\frac{1}{d-2}Tg_{\mu\nu}+\frac{2\Lambda}{d-2}g_{\mu\nu},
\ee
where $R_{\mu\nu}, T_{\mu\nu}$ respectively denote Ricci curvature and energy-momentum tensor, $T$ is the trace of energy-momentum tensor and $\Lambda$ is the cosmological constant.
\vspace*{1mm}

%%%%%%%%%%%%%%%%%%%%%%%%%%%%%%%%%%%%%%%%%
\noindent\textbf{\textit{Theorem 1.}} \emph{
Near horizon of EVH black hole solutions in Einstein gravity coupled with matter fields which have finite and analytic energy momentum tensor at the horizon and $T_{\phi a}=T_{v a}=0$, have a three dimensional locally maximally symmetric part.}
%%%%%%%%%%%%%%%%%%%%%%%%%%%%%%%%%%%%%%%%%

\noindent\textbf{\textit{Proof.}} Recalling that in the Gaussian null coordinates $g_{rr}, g_{ra}$ components of the NHEVH metric ansatz \eqref{NHEVH} are zero, smoothness and analyticity of the energy-momentum tensor at the horizon at $r=0$, which is a generic feature of black hole solutions, implies \cite{Long-version}
\be\label{Rrr-Rra}
R_{rr}=0,\qquad R_{ra}=0.
\ee
These imply that $g_a=0$ and $f_a=\partial_a G/G$. Next, if we also assume that $T_{va}$ and $T_{\phi a}$  vanish for the near horizon EVH geometry, Einstein equations \eqref{Ein-Eq} restrict the form of metric \eqref{NHEVH} to
\bea\label{NHEVH-ansatz-theorem-1}
\hspace{-3mm}ds^{2} = e^{-2K}\left[A_0\rho^2 dv^{2} +2 dv d\rho + \rho^2 d\phi^{2}\right]+\gamma
_{ab} dx^{a}dx^{b},
\eea
where $G=e^{2K}$, $\rho=re^{K}$, $A_0$ is a constant and $\phi$ coordinate in the above is related to $\phi$ in \eqref{NGC-generic} by a  $\phi-cv$ shift for a constant $c$.

One can show through computation of the Riemann curvature and also working out the Killing vectors, that the 3d $v,\rho,\phi$ part of the metric \eqref{NHEVH-ansatz-theorem-1}  is a maximally symmetric space;  for $A_0$ positive, zero and negative, it is respectively locally dS$_3$, flat and AdS$_3$ \cite{Long-version}.

It worths noting  that conditions of Theorem 1 on the energy momentum tensor of matter fields are satisfied in Einstein-Maxwell-Dilaton-$\Lambda$ and gauged or ungauged supergravity theories with U(1) gauge fields, and are not extra conditions.  More detailed analysis is given in \cite{Long-version} and here we sketch the argument. Symmetries of the metric (\ref{NHEVH}) imply scalar fields are only function of $x^a$, therefore $T_{\rho \rho}^{\Phi}=T_{\rho a}^{\Phi}=0, T_{va}^{\Phi} \propto f_a$ and $T_{a\phi}^{\Phi}\propto g_a$. Gauge field potentials of  U(1) gauge fields consistent with symmetries of the metric (\ref{NHEVH}) can be generically written as
\be
A=\rho e dv +\frac{hd\rho}{\rho}+\rho b d\phi+A_a dx^a,
\ee
where $e, h, b$ and $A_a$ are functions of $x^a$. Finiteness of  energy-momentum tensor of the above gauge field at the horizon ($\rho=0$)  implies $b=\partial_ah=0$. Then, equation of motion for gauge fields give $e=0$ and therefore $T_{\rho \rho}^{A}=T_{\rho a}^{A}=0, T_{va}^{A} \propto f_a$ and $T_{a\phi}^{A}\propto g_a$. The same argument can be used for the cosmological constant part of the action. 

Therefore, a general theory of Einstein theory in arbitrary dimensions coupled to scalar and  gauge fields, including all gauged and un-gauged supergravity satisfy conditions of
Theorem 1 and hence, the near horizon geometry of EVH black holes there have generic form \eqref{NHEVH-ansatz-theorem-1}.\ $\mathscr{Q.E.D}$.

\vspace*{1mm}
%%%%%%%%%%%%%%%%%%%%%%%%%%%%%%%%%%%%%%%%
\noindent\textbf{\textit{Theorem 2.}}
\emph{In theories of gravity with matter fields  which besides  the assumptions of Theorem 1, also satisfy strong energy condition, and with non-positive cosmological constant $\Lambda$,  the 3d part of near horizon of an EVH black is  AdS$_3$ for $\Lambda<0$ and for $\Lambda=0$ either it is an AdS$_3$ or the geometry
is a direct product of a locally 3d flat space and a $d\!-\!3$ dimensional part.}
%%%%%%%%%%%%%%%%%%%%%%%%%%%%%%%%%%%%%%%%%

\noindent\textbf{\textit{Proof.}}
The strong energy condition stipulates that
\be
(T_{\mu\nu}-\frac{1}{d-2}\;T g_{\mu\nu})t^{\mu} t^{\nu}\geq 0
\ee
 for every future-oriented timelike vector field $t^{\mu}$.
Eliminating $T_{\mu\nu}$ from the Einstein equations for metric \eqref{NHEVH-ansatz-theorem-1} we arrive at
\be\label{SEC-I}
\nabla^2K-3(\nabla K)^2 -\frac{2\Lambda}{d-2}+2A_0 e^{2K} \leq 0\,,
\ee
where $\nabla^2$ denotes the Laplacian computed with metric $\gamma_{ab}$ and $(\nabla K)^2=\gamma^{ab}\partial_aK\partial_b K$.
Multiplying \eqref{SEC-I} by $e^{-\alpha K}$ with $\alpha \geq 3$ and integrating it on $d-3$ dimensional part, when $\gamma$ has a finite volume we get
$$
\hspace{-3mm}\int_{\gamma_{d-3}}\hspace{-5mm} d^{d-3}x \sqrt{\det\gamma}\ e^{-\alpha K} \bigg[\frac{\alpha-3}{2}(\nabla K)^2-\frac{\Lambda}{d-2}+e^{2K}A_0\bigg] \leq 0.
$$
Therefore, if $\partial_aK\neq0$ then $A_0<0$ for any ${\Lambda}\leq0$ and the near-horizon EVH geometry contains an AdS$_3$ factor. The flat 3d case, $A_0=0$, is only possible when $K=const.$ and $\Lambda=0$, where the warp factor $e^{-K}$ becomes is a constant.  For $\Lambda>0$ cases, the above analysis does not yield a restriction on the sign of $A_0$. \ $\mathscr{Q.E.D}$.

%\begin{flushright} $\square$\end{flushright}
\vspace*{1mm}
%%%%%%%%%%%%%%%%%%%%%%%%%%%%%%%%%%%%%%%%
\noindent\textbf{\textit{Theorem 3.}}
\emph{In theories with non-positive cosmological constant, the 3d part of near horizon of a near-EVH black hole  is either a BTZ black hole or a rotating massive particle on the flat spacetime.}
%%%%%%%%%%%%%%%%%%%%%%%%%%%%%%%%%%%%%%%%

\noindent\textbf{\textit{Proof.}} The near-EVH near-horizon geometries are of the form \eqref{genericNH} with $\epsilon\sim \lambda$. The parameter  $\alpha=\epsilon/\lambda$ ($0 \leq \alpha\lesssim 1$) measures ``out-of-EVH-ness''  and $\alpha=0$ corresponds to the EVH point. We again invoke Einstein equations for determining or restricting the unknown functions in the near-EVH metric. Since these equations should be valid for arbitrary $\alpha$ in the given range, these equations may be expanded in powers of $\alpha$. One then has the zeroth order EVH ($\alpha=0$) results to obtain
\be\begin{split}
&\hspace*{-.15cm}ds^2=e^{-2K}\bigg[-\rho(\rho\tilde{F}+\alpha F^{(1)}) dv^{2}+ 2\rho(\tilde{H}\rho+\alpha H^{(1)}) dv d\phi+\cr&\hspace*{-.25cm}+ 2dv d\rho
+ [(\rho+\alpha R)^2+\alpha^2 J ]d\phi ^{2}  \bigg]
+2\alpha g^{(1)}_a dx^a d\phi  +\gamma_{ab} dx^{a}dx^{b},\nonumber
\end{split}\ee
where $\tilde{F}, \tilde{H}$ are constants by virtue of zeroth $\alpha$ order equations, while $F^{(1)}$ is a constant because it is related to the surface gravity of the near-EVH black hole. The above metric has $d$ more unknown functions $H^{(1)}, R, J$ and $g^{(1)}_a$.
These unknown functions may be determined through the higher order $\alpha$ terms of the equations of motion.

Requiring  $| \partial_\phi|^2=g_{\phi\phi} \geq 0$ everywhere in the spacetime, implies $(\rho+\alpha R)^2$ and $\alpha^2 J$ terms should be non-negative separately and hence $J \geq 0$. Moreover, one would expect that determinant of constant $v$ and $\rho$ sectors should be positive, that is $\det\gamma\cdot [(\rho+\alpha R)^2+\alpha^2 J-e^{2K}\alpha^2\gamma^{ab}g_ag_b]\geq 0$,  where $\gamma^{ab}$ is the inverse of $\gamma_{ab}$. Since $\det\gamma>0$ and that this relation should hold everywhere, we learn that $e^{2K}\gamma^{ab}g_ag_b \leq J$.

Again, smoothness and analyticity of energy-momentum tensor in the near-horizon limit implies $T_{\rho\rho}=T_{\rho a}=0$. These conditions remain true for near-EVH case and therefore we have $R_{\rho\rho}=R_{\rho a}=0$, which in turn yields
\be
J=e^{2K}\gamma^{ab}g^{(1)}_ag^{(1)}_b\,.
\ee
With the above, the metric may be written as
\be\begin{split}
&\hspace*{-.15cm}ds^2=e^{-2K}\bigg[-\rho(\rho\tilde{F}+\alpha F^{(1)}) dv^{2}+ 2\rho(\tilde{H}\rho+\alpha H^{(1)}) dv d\phi+\cr&\hspace*{-.25cm}+ 2dv d\rho
+ (\rho+\alpha R)^2d\phi ^{2}\bigg]+\gamma_{ab}(dx^{a}+\alpha{\hat g}^a d\phi)(dx^{b}+\alpha{\hat g}^b d\phi)\;,\nonumber
\end{split}\ee
where $\hat{g}^a =\gamma^{ab}g^{(1)}_b$. Analysis of equations of motion and in particular with the $T_{va}=0, T_{\phi a}=0$, does not yield $\hat{g}^a=0$, while they imply $R,\ H^{(1)}$ are constants, if we assume $\hat{g}^a=0$ \cite{Long-version}. The $\hat{g}^a=0$ assumption is equivalent to $\partial_\phi$ be a hypersurface orthogonal  Killing vector on the horizon of the original EVH black hole; i.e. at codimension two constant $v$ and $r=0$ surfaces, $\partial_\phi$ is transverse to the constant $\phi$ surfaces \cite{Footnote-near-EVH}. With the above assumptions, we obtain a metric with five constants, one of which can be removed by a coordinate transformation $\phi\to \phi+ c v$, with a constant $c$, and the unknown functions $K$, $\gamma_{ab}$.

The 3d $\rho,v,\phi$ part is not a maximally symmetric space, unless the constants are related in a specific way. Such relations may come from components of the Einstein equations along the 3d part. Explicit computations of the energy momentum tensor for generic tensor (Einstein)-vector (Maxwell field)-scalar theories \cite{Long-version} reveal that the energy momentum along the 3d part is proportional to its metric, implying that
\be\label{flat-condition}
H^{(1)}=2\tilde H R\,,\qquad F^{(1)}=2\tilde F R.
\ee
With the above, the 3d part of the near-EVH metric becomes a locally constant curvature space.

As in the EVH case, if the matter fields satisfy strong energy condition, we deal with two options:\\
$\bullet$  $A_0=-(\tilde F+\tilde H^2)<0$, then we have a locally AdS$_3$ space, with metric
\bea\label{NH-Near-EVH-AdS-case}
&& ds^{2} = e^{-2K}\!\bigg[-\tilde F\rho (\rho+2\alpha  R) dv^{2}+ 2\tilde{H} \rho(\rho+2\alpha  R) dv d\phi\nonumber
\\
&&\hspace{10mm} + (\rho+ \alpha R)^2d\phi ^{2} +2 dv d\rho \bigg]\!
+\gamma_{ab}dx^{a}dx^{b},
\eea
which its 3d part denotes a BTZ geometry \cite{BTZ}, written in Gaussian null coordinates, with inner and outer horizon radii $r_\pm$ and AdS$_3$ radius $\ell$
\be
\ell^2=-\frac{1}{A_0}\,,\qquad r_+=\alpha R\,,\qquad \tilde H=\frac{r_-}{\ell r_+}\,.
\ee
We note that if the $\phi$ direction in the original EVH black hole (before taking the near horizon limit) was ranging over $[0,2\pi]$, after taing the near horizon limit \eqref{NH-limit} the $\phi$ direction will be ranging over $[0,2\pi\lambda]$. This geometry is hence called ``pinching BTZ'' \cite{BTZ-EVH}.\\
$\bullet$  For $A_0=0$ after the shift $\rho\to \rho-\alpha R$ and $\phi\to\phi-\tilde Hv$, and then rescaling $v, \phi$ and $\rho$, metric takes the form
\be\label{NH-Near-EVH-flat-case}
 ds^{2} = e^{-2K} \big[ dv^{2}+ \frac{2}{\tilde{H}}dv d\phi+ \rho^2 d\phi ^{2} +2 dv d\rho \big] +\gamma_{ab}dx^{a}dx^{b},
\ee
where the $\phi$ coordinate is ranging over $[0,2\pi \alpha\tilde H R\lambda]$. The 3d part of metric is locally flat and represents a particle of a given mass and spin proportional to $\tilde H$ \cite{DJT}.\  $\mathscr{Q.E.D}$.
\vskip 1.5mm

\textbf{{Concluding remarks.}}
%%%%%%%%%%%%%%%%%%%%%%%%%%%%%%%%%%%%
In this work we proved three  theorems regarding near horizon limit of (near) Extremal Vanishing Horizon black hole solutions. Our results are interesting and powerful because they apply to quite generic gravity theories in diverse dimensions and recalling that in the context of gravity solutions we do not usually have such theorems. In a sense our reasoning is close, and our results are complementary, to similar analysis for extremal black holes \cite{KL-papers,KL-review}. Our theorems state that for theories obeying strong energy condition we generically get an AdS$_3$ factor in our NHEVH geometry. While  the possibility of 3d flat space is not ruled out by our theorems, we do not know any explicit example which actually realizes this possibility. It would hence be interesting to explore if this possibility can be ruled out through some other properties (e.g. other energy conditions) of the matter fields.

Our theorems are not uniqueness or classification theorems, neither for the EVH black holes nor for their near horizon limits; they uncover interesting, generic features of NHEVH geometries. However,  our theorems once considered together with analysis of \cite{EVH/CFT,SO22}, provide a classification and uniqueness for four and five dimensional NHEVH solutions to Einstein-Maxwell-Dialton theories.

An important point to keep in mind regarding the AdS$_3$ throat we get is that, as seen from \eqref{NH-limit}, the $\phi$ direction is a ``pinching'' direction \cite{BTZ-EVH}, it is ranging over $\phi\in [0,2\pi\lambda]$, if the $\phi$ direction in the original black hole had a $[0,2\pi]$ range. Then, one should note this fact if based on this pinching AdS$_3$ one wants to put forward an EVH/CFT correspondence \cite{EVH/CFT,EVH-examples}.
In this respect and recalling our near-EVH theorem (Theorem 3), the EVH case is interesting because, unlike the extremal case \cite{No-dynamics},  it allows for ``excitation'' and nontrivial dynamics about the NHEVH geometry.
In particular, it would be interesting to explore further how the laws of black hole thermodynamics after and before the near horizon limit for near-EVH black holes are related. First steps in this direction is taken in \cite{First-law}.
\begin{acknowledgements}

MMSHJ and SS  would like to thank Allameh Tabatabaii Prize Grant of Boniad Melli Nokhbegan of Iran. MMSHJ, SS and MHV would like to thank the ICTP network project NET-68.

\end{acknowledgements}


\begin{thebibliography}{99}
\bibitem{Footnote-near-EVH}
The $\hat g^a=0$ assumption is not relevant for 4d cases, as discussed in \cite{EVH/CFT}, and it remains vaild for all the higher dimensonal examples explored
in \cite{EVH-examples}.
\end{thebibliography}
\end{document}